\documentclass[12pt]{article}
\usepackage{titlesec}
\usepackage{ntheorem}
\usepackage{tikz}
\titlespacing*{\section}
{0pt} 
{1pt} 
{1pt}  

\titlespacing*{\section}
{0pt} 
{1pt} 
{1pt}  

\titlespacing*{\subsubsection}
{0pt} 
{0pt} 
{0pt}  
\colorlet{highlightrow}{blue!12}
\colorlet{highlightcell}{red!35}

\def\xa{0.72} \def\xb{0.64}   
\def\ya{0.38} \def\yb{0.30}   
 
\newcommand{\iagrid}{%
  \draw[thick] (0,0) rectangle (1,1);
  \draw[dashed] (\xa,\xb) -- (\xa,0);
  \draw[dashed] (\xa,\xb) -- (0,\xb);
  \draw (\ya,0) -- (\ya,\xb);
  \draw (0,\yb) -- (\xa,\yb);
  \node[below] at (0,0) {$0$};
  \node[below] at (1,0) {$1$};
  \node[left]  at (0,1) {$1$};
  \node[below] at (\xa,0) {$x_1$};
  \node[below] at (\ya,0) {$y_1$};
  \node[left]  at (0,\xb) {$x_2$};
  \node[left]  at (0,\yb) {$y_2$};
}

 \usepackage{booktabs}
\usepackage{colortbl}
\usepackage{xcolor}
\usepackage{array}
\usepackage{diagbox}

\definecolor{barone}{HTML}{4C72B0}
\definecolor{bartwo}{HTML}{DD8452}
\newcommand{\splitcell}[2]{\textcolor{barone}{#1}\textcolor{gray}{/}\textcolor{bartwo}{#2}}

\newcolumntype{C}{>{\centering\arraybackslash}p{1.25cm}}
\newcolumntype{S}{>{\centering\arraybackslash}p{0.5cm}}
\newcolumntype{D}{>{\centering\arraybackslash}p{0.55cm}}
\newcommand{\rs}{\rule{0pt}{3.7mm}}

\newcommand{\iabox}[3]{%
  \begin{tikzpicture}[scale=2.3, every node/.style={font=\footnotesize}]
    \fill[gray!30] (0,0) rectangle (#1,#2);
    \iagrid
    \fill (#1,#2) circle (0.5pt);
    \node at ({\ya/2},{\yb/2}) {$#3$};
  \end{tikzpicture}%
}

 \usepackage{pdfpages}
\usepackage{colortbl}
\usepackage{booktabs}
\usepackage{bbm}
\usepackage[normalem]{ulem}

\newcommand{\R}{\mathbb{R}}

\newcommand{\N}{\mathbb{N}}

\newcommand{\id}{\mathds{1}}

\renewcommand{\epsilon}{\varepsilon}

\usepackage{charter}
\usepackage{pgf}
\usepackage{tikz}
\usetikzlibrary{arrows,automata}
\usepackage{booktabs} 
\usepackage{verbatim}
\usepackage{subcaption}

\usepackage{xcolor}

\usepackage{booktabs}
\usepackage{bbm}

\usepackage{natbib}

\colorlet{drkblue}{blue!61.8!black}

\usepackage[bookmarks=false]{hyperref}
\hypersetup{
     pdfborder = {0 0 0},
  colorlinks=true,
 citecolor=drkblue,
linkcolor=drkblue,
urlcolor=drkblue,
pdfstartview={XYZ null null 0.80}
}

\usepackage[hyperpageref]{backref}

\usepackage{amsmath}

\usepackage[left=2.5cm,top=3.5cm,bottom=3.5cm,right=3cm]{geometry}

\usetikzlibrary{math}

\usepackage{amssymb,amsmath,xcolor}

\newtheorem{theorem}{Theorem}
\newtheorem{proposition}{Proposition}

\theorembodyfont{\rmfamily}
\theorembodyfont{\rmfamily}
\newenvironment{proof}[1][Proof:]{\noindent\textbf{#1} }

\theorembodyfont{\rmfamily}
\newdimen\dummy
\dummy=\oddsidemargin
\usepackage{graphicx}
\usepackage{amsfonts}
\usepackage{amsmath}
\usepackage{amssymb}
\usepackage{mathabx}
\usepackage{url}
\usetikzlibrary{tikzmark}
\usepackage{graphicx}
\usepackage{tikz,lipsum,lmodern}
\usepackage{dsfont}
\usepackage{fancyhdr}
\usepackage{indentfirst}
\usepackage{enumitem}
\usepackage[normalem]{ulem}
\usepackage{mathrsfs}
\usepackage{multirow}
\usepackage{diagbox}

\title{Invariant Modeling for Joint Distributions\thanks{We thank the audience of BRIC X, the Herv\'{e} Moulin 75th birthday Conference, the Shanghai Microeconomics Workshop 2025,  the Science of Decision Making (SDM) Conference, and the theory seminar participants at UC Berkeley, Cornell, Caltech, Princeton, and Rochester. Special thanks go to Tri Phu Vu, who provided valuable assistance in carefully reviewing the proofs for this manuscript. All remaining errors are ours.}}

 \author{\hspace*{0cm} Christopher P. Chambers\thanks{Department of Economics, Georgetown University. 
 E-mail: \texttt{Christopher.Chambers@georgetown.edu}} \and Yusufcan Masatlioglu\thanks{Department of Economics, University of Maryland. 
 E-mail: \texttt{yusufcan@umd.edu}} \and Ruodu Wang\thanks{Department of Statistics and Actuarial Science, University of Waterloo. 
 E-mail: \texttt{wang@uwaterloo.ca}  Ruodu Wang is supported by the Natural Sciences and Engineering Research Council of Canada (CRC-2022-00141, RGPIN-2024-03728).}.  }
 \date{\today}

\begin{document}

\maketitle

\begin{abstract}
A common theme underlying many problems in statistics and economics involves the determination of a systematic method of selecting a joint distribution consistent with a specified list of categorical marginals, some of which have an ordinal structure.  We propose guidance in narrowing down the set of possible methods by introducing  Invariant Aggregation (IA), a natural property that requires merging adjacent categories in one marginal not to alter the joint distribution over unaffected values. We prove that a model satisfies IA if and only if it is a copula model. This characterization i) ensures robustness against data manipulation and survey design, and ii) allows seamless incorporation of new variables. Our results provide both theoretical clarity and practical safeguards for inference under marginal constraints.


\end{abstract}

\maketitle

\section{Introduction}
Recovering a joint distribution from its marginal distributions is a fundamental problem in statistics, economics, machine learning, and related fields. However, marginal distributions alone are generally insufficient to uniquely determine the joint distribution—infinitely many joint distributions may be consistent with the same set of marginals. This ambiguity grows with dimensionality, as the number of free parameters in the joint distribution increases exponentially with the number of random variables in discrete statistical models.

To resolve this indeterminacy, additional structure must be imposed--such as assumptions about dependence patterns, joint moments, or other inter-variable relationships. A fundamental approach is to introduce a functional mapping, denoted by $f$, that takes a set of {discrete} marginal distributions and returns a unique joint distribution consistent with them. We refer to such a mapping $f$ as a model. Each model imposes certain structural assumptions, such as positive dependence, negative dependence, and independence. 

In this paper, we propose and analyze an intuitive structural property that a model $f$ may satisfy: aggregating two adjacent values in the support of one marginal should not affect the joint distribution over the remaining, unaltered values. The determination of adjacency requires us to specify an order on each marginal. Our approach is motivated by examples in economics, but the ideas here may be of interest in other fields such as political science, statistics, or psychology. 

To illustrate, consider electricity and water consumption over a fixed period of time in a neighborhood.  Each consumer in the neighborhood chooses some amount of water, and some amount of electricity.  Water and electricity consumption are each naturally modeled as real-valued, but we suppose that each of these real-valued consumption levels is partitioned into bins: low, medium, and high ($L$, $M$, and $H$).  Each bin represents an interval of possible real-valued consumption levels.  Importantly, a consumer whose water choice lies in bin $L$ consumes less water than one whose choice lies in bin $M$; we write $L<M<H$.  The economist observes the empirical distribution of choices from bins $L$, $M$ and $H$ for both water consumption across all consumers, and for electricity consumption.  We refer to these as \emph{marginals}.

Even though each consumer chooses both water and electricity consumption, we assume that these joint choices are unobserved.  That is, we imagine data comes from different sources, leaving the joint distribution of water and electricity consumption unknown.  We seek to study methods of selecting joint distributions.  As a modeler or policymaker, we may hypothesize a positive association—for instance, we may conjecture that higher electricity usage may tend to coincide with higher water usage. Given the marginals, our model produces a corresponding joint distribution.  But many other hypotheses are possible.

Now, imagine that due to {a different survey design}, the medium and high electricity categories are merged (written $MH$), while the water usage categories remain unchanged. The marginal distribution for electricity becomes coarser, and the model generates a new joint distribution. Crucially, since the low category is unaffected, we expect the joint probabilities involving this category recommended by our selection to remain unchanged. Our proposed property, {called Invariant Aggregation (IA)}, ensures this invariance, enhancing robustness and interpretability. IA captures the idea that predictions of a model would not contradict themselves in similar situations. In other words, the prediction of a model is invariant to the small changes in response categories, which is a desirable property in categorical data analysis \citep{agresti2010analysis,johnson2006ordinal}.

Violating IA would render social policies sensitive to arbitrary data-reporting choices—especially problematic when reporting methods vary across contexts. Worse, it could allow unscrupulous data collectors to manipulate outcomes by altering data presentation. Many, but not all, models satisfy IA. A trivial example of a model satisfying IA is the independence model, which assumes statistical independence across variables, constructing the joint distribution via multiplication of marginals. However, this is not the only such model. Our main result provides a complete characterization of all IA-compliant models: they are precisely those defined by copulas, a well-studied class of statistical objects. Indeed, for statisticians, to get from marginal distributions {on the real line} to a joint distribution, copulas come to mind first. Sklar's theorem (see \citealp {nelsen2006}) states that a joint distribution on a Euclidean space can be described by its marginal distributions and a copula.\footnote{Copulas have gathered great attention from both  statistical theory  \citep{genest1993statistical, joe2014dependence} and many application areas, including finance \citep{li1999default, mcneil2015quantitative},  econometrics \citep{beare2010copulas, fan2023vector}, stochastic choice theory \citep{chambers2025ordered}, bargaining \citep{bastianello2019}, and behavioral game theory \citep{frick2022dispersed}, in addition to the natural sciences.} Our task is, however, different: in our setting, both the marginal distributions and the joint distribution vary (and they are categorical), while keeping the invariance property.

Finally, IA has an important implication for introducing new variables. If the $i$-th dimension initially contains no data, increasing its support cardinality is akin to introducing new data. Critically, any IA model accommodates such extensions without altering the joint distribution over existing dimensions.  

Other sections of the paper are organized as follows. Section \ref{sec:model} introduces the notation and definitions of IA.  Section \ref{sec:result} introduces and proves the main result of the paper. Section \ref{sec:cat} investigates an environment where the order structure plays no role, and bins can be arbitrarily permuted. We show that a model satisfying IA in this case must be associated with the independence copula.  Section \ref{sec:add} describes two additional results: first, a result claiming that any model satisfying IA which has the property that identical marginals are mapped to the perfectly correlated joint distribution must be associated with the Fr\'echet upper bound copula,
and second, a result investigating the introduction of new variables.  In this result, we hypothesize that the IA model should be symmetric, in the sense that all commodities are treated the same.  We require that the introduction of new variables can be accommodated with a new symmetric model satisfying IA, and leverage de Finetti's Theorem to show that this property characterizes a simple class of copula models.  Finally, Section \ref{sec:conclusion} discusses related literature and concludes.

\section{Model}\label{sec:model}

We have in mind the following problem. There are several categorical
random variables.  To represent these random variables, fix $n\in\mathbb{N}$, and assume given an ordered list of $n$ finite sets $(A_1,\ldots,A_n)$, each of which is endowed with an ordinal ranking $<_i$ over $A_i$. Although the variables are categorical in nature, the underlying implicit structure we envision consists of a real-valued random variable that has been partitioned into a finite number of bins, each of which represents an interval. For example, political positions in the ``spatial model'' of politics are often modeled as a unidimensional spectrum (\emph{e.g.}~Liberal, Moderate, Conservative) \citep{downs1957}, and the importance of global warming could be categorized into five bins (extremely, very, somewhat, not too, not at all important) \citep{Chryst2018}.  

For each set $A_i$, a probability distribution $p_i$ is given over $A_i$.  Our goal is to find a systematic method of constructing a joint probability distribution $p$ on $\prod_{i=1}^n A_i$ whose marginals coincide with each $p_i$.

Our framework is ordinal, so the actual names of the elements of each $A_i$ are not relevant.  Instead, only the cardinality and order structure of each $A_i$ matter.  
Fix $n\in \mathbb N$ and let $N=[n]$.  
Therefore, the variable $n$ represents the number of sets $A_i$ for which we anticipate receiving data.  For example, consider the example discussed in the introduction, where a distribution of water consumption is specified, and a distribution of electricity consumption is specified.  In this example, we can imagine that $A_1$ represents possible water consumption levels, and $A_2$ represents possible electricity consumption levels.  

As is standard, for any countable set $A$, let $\Delta(A)$ be the set of all probability distributions on $(A,2^A)$.  For a probability distribution $q$ on $\R^n$, we denote by $[q]_i$ its $i$th marginal distribution for $i\in N$.\footnote{Similarly for a distribution on any subset of $\R^n$.} Distributions are identified with their probability mass functions; that is, we write $q(a)=q(\{a\})$ for $a\in A$.

For any $k\in\mathbb{N}$, let $[k]=\{1,\ldots,k\}$. Then, for $z=(z_1,\dots,z_n)\in\mathbb{N}^n$, 
denote by $[z]=\prod_{i=1}^n [z_i]$.  For each $i\in N$, each $[z_i]$ reflects the finite number of bins ($z_i$ bins) from which individuals may choose, or the cardinality of $A_i$.  Return to our water and electricity example.  If $A_1$ reflects water consumption, then $z_1$ is simply a label of the highest bin (our introductory example featured labels of the form $L<M<H$; here the integers $1<2<3$ are placeholders for these labels.  In this example then, $z_1=3$).  Thus, in the economics example, an element of $[z_i]$ constitutes a choice of the $i$-th good.  In this case, a member $s$ of $[z]$ reflects a hypothetical or possible \emph{joint} consumer choice:  a possible consumer whose water and electricity consumption fall in bins $s_1$ and $s_2$, respectively.

Our setup assumes that the actual contents of each bin are irrelevant.  We only use the ordinal ranking across bins.  We think this specification is useful: it keeps our analysis both tractable and applicable in a broad variety of situations.  A finite list of ordered bins in our model is always written as $\{1,\ldots,z_i\}$, for some $z_i\in\mathbb{N}$, without reference to the names of the alternatives in the bins.  Returning to the $L<M<H$ example from the introduction, observe that $\{L,M,H\}$ is a three-element set, so we would set $z_i=3$ in this case.  In merging $L$ and $M$, we obtain $LM<H$, where $\{LM,H\}$ is a two element set, so $z_i=2$.  Similarly, merging $M$ and $H$ results in a two element set $L<MH$, with $z_i=2$.  Our framework therefore does not include a language for distinguishing the ordered set $\{L,M\}<H$ from $L<\{M,H\}$.  We discuss this further after the statement of the main result.

Next, define $$\Delta^n(z)=\prod_{i=1}^n \Delta([z_i])= \{(p_1,\dots,p_n): p_i\in \Delta([z_i]), ~i\in N\}.$$  For a given $z$, any member of $\Delta^n(z)$ constitutes the input of the model. In practical settings, the input could consist of the empirical distributions of variables of interest.   It is a list of probability distributions, consisting of a probability distribution over $[z_i]$ for each $i\in N$. 

We need a method for describing how data can change as $z$ changes---this was our primary motivation.  The bins into which choices can be put can be refined further.  To this end, let  $$\Delta^n_{\rm ma} =\bigcup_{z\in \N^n} \Delta^n(z)
\mbox{~~and~~} \Delta^n_{\rm jo} =\bigcup_{z\in \N^n}\Delta\left ( [z]\right).$$  
{The set} $\Delta^n_{\rm ma}$ lists all possible datasets we may ever possibly encounter as $z$ varies.  On the other hand, members of $\Delta^n_{\rm jo}$ reflect our possible \emph{models} of data.  For any $[z]$, a member of $\Delta\left([z]\right)$ is a joint distribution over $\prod_{i\in N}[z_i]$; $\Delta^n_{\rm jo}$ collects all of these as $z$ varies.  

Consider the following running example. There are two random variables ($A_1$ and $A_2$ are water and electricity consumptions). Hence $n=2$. There are three categories in $A_1$ and four categories in $A_2$, hence $(z_1,z_2)=(3,4)$.  Then data comes in the form of a pair of distributions:  a distribution $p_1$ over $\{1,2,3\}$ and a distribution $p_2$ over $\{1,2,3,4\}$.  Our aim is to construct a joint distribution on $\{1,2,3\}\times\{1,2,3,4\}$ whose marginals are $p_1$ and $p_2$ respectively.  We seek a method for doing this no matter what our choice of $(z_1,z_2)$ is and no matter what the marginals may be.  The left part of Figure~\ref{fig:independence} illustrates $(z_1,z_2)=(3,4)$.  The top row summarizes $z_2$, and the leftmost column represents $z_1$. This idea motivates the following definition, which is our main object of study.

For vectors  $z,s\in \mathbb N^n$ and $p\in \Delta^n_{\rm ma}$, as well as similar quantities, we tacitly denote their components by $(z_1,\dots,z_n)$, $(s_1,\dots,s_n)$, and $(p_1,\dots,p_n)$, respectively. 
For notational convenience, in what follows, when $p_1,\dots,p_n$ appear, their domains are   $[z_1],\dots,[z_n]$, unless otherwise specified.

Let $f$ be a mapping from $\Delta^n_{\rm ma}$ to $\Delta^n_{\rm jo} $
satisfying
$[f(p)]_i = p_i$ for $p\in  \Delta^n_{\rm ma}$ and $i\in N$. 
Such a mapping $f$ is called 
 a \emph{model}. Perhaps the simplest example of a specific model is the \emph{independence model}, given by
 $$
 f(p)(s)=\prod_{i\in N} p_i(s_i),~~~s\in [z] 
 $$  

The phrase ``model'' is intended to be understood as a method for understanding or organizing data:  it is not intended to be understood as a falsifiable behavioral model.  The function $f$ constitutes a ``modeling assumption'' in the sense that the rule for choosing the joint distribution is specified independently of the data themselves.  In our context, we are typically thinking of observing commodity or goods choice with no observed prices or price variation, so it is difficult to think of how one should build a more structural behavioral model in any case.  

Return to the preceding example, whereby $(z_1,z_2)=(3,4)$.  We will show how to apply the independence model to data.  Suppose that $p_1=(1/2,1/4,1/4)$ and $p_2=(1/4,1/4,1/4,1/4)$.  If $f$ is the independence model, then for any pair $(s_1,s_2)\in [z]$, $f(p)(s_1,s_2)=p_1(s_1)p_2(s_2)$.  So, for example, $f(p)(1,2)=1/8$.  Observe that $f(p)$ is a probability distribution over $[z]$ and further that the marginal of $f(p)$ in each dimension coincides with the original data.  Figure~\ref{fig:independence} illustrates the application of the independence model.  The bottom row summarizes the marginal probabilities associated with $p_2$, and the rightmost column the marginal probabilities associated with $p_1$.  So, we can find the joint probability of $s=(2,3)$, for example, by multiplying these values:  it is $1/16$. 

 


\begin{figure}[ht]
\centering
\renewcommand{\arraystretch}{0.8}
\begin{minipage}[t]{0.435\textwidth}
\centering
{\footnotesize Before refinement}\\[.2em]
\begin{tabular}{r|cccc|l}
\toprule
 & $1$ & $2$ & $3$ & $4$ & $p_1$ \\
\midrule
$1$ & $1/8$ & $1/8$ & $1/8$ & $1/8$ & $1/2$ \\
$2$ & $1/16$ & $1/16$ & $1/16$ & $1/16$ & $1/4$ \\
\rowcolor{highlightrow}
$3$ & $1/16$ & $1/16$ & $1/16$ & $1/16$ & $1/4$ \\
\\
\midrule
$p_2$ & $1/4$ & $1/4$ & $1/4$ & $1/4$ & \\
\bottomrule
\end{tabular}
\end{minipage}
\hfill
\raisebox{-2.4cm}{ \Large $\Longrightarrow$}
\hfill
\begin{minipage}[t]{0.435\textwidth}
\centering
{\footnotesize After refinement}\\[.2em]
\begin{tabular}{r|cccc|l}
\toprule
 & $1$ & $2$ & $3$ & $4$ & $p_1'$ \\
\midrule
$1$ & $1/8$ & $1/8$ & $1/8$ & $1/8$ & $1/2$ \\
$2$ & $1/16$ & $1/16$ & $1/16$ & $1/16$ & $1/4$ \\
\rowcolor{highlightrow}
$3$ & $1/32$ & $1/32$ & $1/32$ & $1/32$ & $1/8$ \\
\rowcolor{highlightrow}
$4$ & $1/32$ & $1/32$ & $1/32$ & $1/32$ & $1/8$ 
\\
\midrule
$p_2$ & $1/4$ & $1/4$ & $1/4$ & $1/4$ & \\
\bottomrule
\end{tabular}
\end{minipage}
\caption{\footnotesize Illustration of the independent copula before and after refinement.  The independent copula applied to marginals $p_1$ (rows) and $p_2$ (columns): each cell equals $p_1(i)\cdot p_2(j)$. Refining category $3$ into categories $3$ and $4$, holding $p_2$ fixed, leaves rows $1$ and $2$ completely unchanged and splits row $3$'s mass in exact proportion to the new categories' shares.}
\label{fig:independence}
\end{figure}

Our notion of a model needs to specify what happens for more refined data.  Let us now imagine $(z_1',z_2)=(4,4)$. For example, $A_1$ (the water consumption) is provided with finer categories.  Consider $p_1'\in\Delta([z_1'])$ given by $(1/2,1/4,1/8,1/8)$.  This data is consistent with the idea that the last bin in $[z_1]$ splits into two bins while the other two bins do not change:  the probability of the initial two bins remains the same while $p_1'(3)+p_1'(4)=p_1(3)$.  Thus, $p'_1$ is consistent with the idea that the data described by $p_1$ has become more refined.  Now, let us apply the independence model to $(p_1',p_2)$.  For example, $f(p_1',p_2)(3,2)=p_1'(3)p_2(2)=1/32$ and $f(p_1',p_2)(4,2)=p_1'(4)p_2(2)=1/32$.  


Observe that $f(p_1',p_2)(1,2)=1/8=f(p_1,p_2)(1,2)$.  Indeed, the entries in the first two rows did not change. That is, the recommendations made by {the independence model} for the coarse and the fine datasets are consistent.

To formalize the idea behind the preceding example, we need to introduce some new notation.  Specifically, we need to describe a notation that allows us to merge two bins into one.

Take any $w\in \{2,3,\dots\}$ and $q \in \Delta([w])$.
 For  $j\in [w-1]$,  let $q^{j} \in \Delta ([w-1])$ be given by, for all $t \in [w-1]$, $q^{j}(t) = q(t)\id_{\{t\le j\}} +q(t+1)\id_{\{t\ge j\}}.$ Then, given $q$, the measure $q^j$ is the measure that results from ``collapsing'' categories $j$ and $j+1$.

When $q^j$ appears for a probability $q\in \Delta([w])$, we always assume  $j\in [w-1]$.
For $i\in N$, $e^i$ is defined as the unit vector $(0,\dots,0,1,0,\dots,0) \in \R^n$
where $1$ is at the $i$th position. 
For any distribution $q$, we  write $F_q$ for the cumulative distribution function (cdf) of $q$.\footnote{For multidimensional distributions, this is defined as $F_q(s)=q(\{t:t_i\leq s_i\mbox{ for all }i\in N\})$.}
Let $T_i^j$ be the mapping that maps $p$ to 
$(p_1,\dots,p_{i-1}, p_i^j,p_{i+1},\dots, p_n)$, for all $i\in [n]$ and $j\in [z_i-1]$.

Thus, $T_i^j(p)$ is simply a profile of marginal choices where, for set $i$, alternatives $j$ and $j+1$ are collapsed into one.

Our primary axiom can now be expressed as follows.  It requires that when merging two adjacent alternatives into one, the model should respond consistently.  

\textbf{Invariant Aggregation (IA):} For all   $p\in \Delta^n_{\rm ma} $, all $i\in N$,  all $j\in[z_i-1]$ and all $s\in[z-e^i]$,  
\begin{equation}
\label{eq:condition-main} 
 f\left(T_i^j(p)\right) (s) 
 = f(p)(s) \id_{\{s_i\le j\}} + f(p)(s+e^{i})\id_{\{s_i\ge j\}}
\end{equation}

Observe that IA imposes no actual restriction on how a model behaves for any particular list of marginals, but rather establishes how a model behaves across lists of marginals.  This type of property is common enough in economics that it has a name (derived from social choice); it is termed a \emph{multi-profile} axiom.  Many of the most well-known characterizations in economics are based on this type of axiom.  For example, Samuelson's classical treatment of what came to be called the weak axiom of revealed preference does not restrict demand at any given budget, but rather specifies how demand must change as budgets change \citep{samuelson1938,rose1958}.
 

Importantly, this axiom is not vacuously satisfied: many natural constructions violate it. We illustrate this with the \emph{maximal coupling}, a standard object in probability theory designed to maximize the agreement between two random variables. For two marginal distributions $p_1$ and $p_2$ with supports of size $z_1$ and $z_2$, respectively, a maximal coupling $f(p)$ maximizes the probability of the diagonal event $\{(s,s):s\leq\min\{z_1,z_2\}\}$. 

Returning to our example, we apply the maximal coupling to the same data as before. Figure~\ref{fig:maxcoupling} displays the resulting joint distribution $f(p)$ before and after refinement. : the left panel corresponds to $(p_1,p_2)$, with $(z_1,z_2)=(3,4)$; the right panel corresponds to $(p_1',p_2)$ after refining category $3$ into categories $3$ and $4$, so that $(z_1,z_2)=(4,4)$. In both panels, the diagonal entries $f(p)(s,s)$ are as large as the marginals allow.

\begin{figure}[ht]
\centering
\renewcommand{\arraystretch}{0.8}
\begin{minipage}[t]{0.435\textwidth}
\centering
{\footnotesize Before refinement}\\[.2em]
\begin{tabular}{r|cccc|l}
\toprule
 & $1$ & $2$ & $3$ & $4$ & $p_1$ \\
\midrule
{$1$} & $1/4$ & $0$ & \cellcolor{highlightcell}$0$ & \cellcolor{highlightcell}$1/4$ & {$1/2$}\\
$2$ & $0$ & $1/4$ & $0$ & $0$ & {$1/4$}\\
\rowcolor{highlightrow}
$3$ & $0$ & $0$ & $1/4$ & $0$ & {$1/4$}\\ 
\\
\midrule
$p_2$ & $1/4$ & $1/4$ & $1/4$ & $1/4$ & \\
\bottomrule
\end{tabular}
\end{minipage}
\hfill
\raisebox{-2.4cm}{ \Large $\Longrightarrow$}
\hfill
\begin{minipage}[t]{0.435\textwidth}
\centering
{\footnotesize After refinement}\\[.2em]
\begin{tabular}{r|cccc|l}
\toprule
 & $1$ & $2$ & $3$ & $4$ & $p_1'$ \\
\midrule
{$1$} & $1/4$ & $0$ & \cellcolor{highlightcell}$1/8$ & \cellcolor{highlightcell}$1/8$ & {$1/2$}\\
$2$ & $0$ & $1/4$ & $0$ & $0$ & {$1/4$}\\
\rowcolor{highlightrow}
$3$ & $0$ & $0$ & $1/8$ & $0$ & {$1/8$}\\
\rowcolor{highlightrow}
$4$ & $0$ & $0$ & $0$ & $1/8$ & {$1/8$}\\
\midrule
$p_2$ & $1/4$ & $1/4$ & $1/4$ & $1/4$ & \\
\bottomrule
\end{tabular}
\end{minipage}
\caption{\footnotesize Illustration of the maximal coupling before and after refinement. Refining category $3$ into categories $3$ and $4$. Even though category 1 is not refined, the cells highlighted in red have changed in row $1$. Hence, IA is violated.}
\label{fig:maxcoupling}
\end{figure}



Observe that IA is violated. Were it satisfied, the probability of $(1,4)$ would need to remain the same across the two environments---but it falls from $1/4$ to $1/8$, even though only category $3$ was refined.

{As a final ingredient, we need the notion of copulas}. A \emph{copula} is a function $C$  from $[0,1]^n$ to $[0,1]$ 
that is a cumulative distribution function for a distribution with standard uniform (i.e., uniform on $[0,1]$) marginals.  So, we may conclude that $C$ is both nondecreasing and continuous, and satisfies other properties, such as $C(1,\ldots,1,x_,1,\ldots,1)=x$, for example.

There are many copulas; a good introduction to the theory is \cite{nelsen2006}.  Two classical ones are as follows:

\begin{itemize}
\item The \emph{independence copula} is given by $C(x_1,\ldots,x_n)=\prod_{i=1}^n x_i$.
\item The \emph{Fr\'echet-Hoeffding upper-bound} is defined as $C(x_1,\ldots,x_n)=\min_{i}x_i$.  
\end{itemize}

These two copulas happen to be described by symmetric functions (permuting the coordinates does not change the value of the copula).  This need not be the case.  For a matrix $\Sigma\in\mathbb{R}^{n\times n}$ corresponding to the correlation coefficients of a multivariate normal, the \emph{Gaussian copula} is given by $C(x_1,\ldots,x_n)=\Phi_{\Sigma}(\Phi^{-1}(x_1),\ldots,\Phi^{-1}(x_n))$, where $\Phi$ is the cdf of a standard normal and $\Phi_{\Sigma}$ is the joint cdf of a mean-zero normal with correlation matrix $\Sigma$.  The independence copula is the special case where $\Sigma$ is the identity matrix.

A \emph{copula model} is a model $f$ for which there is a copula $C:[0,1]^n\rightarrow [0,1]$ such that
\begin{equation}\label{eq:cdf}F_{f(p)}(s) =C(F_{p_1}(s_1),\dots,F_{p_n}(s_n))\end{equation}
for all $p\in \Delta^n_{\rm ma}$ and $s\in \R^n$.  The left-hand side of~\eqref{eq:cdf} reports the cumulative probability  assigned by $f(p)$ to $s$, that is, the probability of $\{s':s'_i \leq s_i\mbox{ for all }i\}$.  As in the standard single-dimensional case, this object uniquely pins down $f(p)$, through an inclusion-exclusion principle.  The right-hand side applies the copula $C$ to the cumulative probabilities of the marginals $p_i$. If $C$ is the independence copula, the corresponding copula model is the \emph{independence model}.

We now demonstrate what happens when the Fr\'echet--Hoeffding upper bound $M$ is applied to the same data. Figure~\ref{fig:mincopula} displays the resulting joint distribution: the left panel corresponds to $(p_1,p_2)$, with $(z_1,z_2)=(3,4)$; the right panel corresponds to $(p_1',p_2)$ after refining category $3$ into categories $3$ and $4$, so that $(z_1,z_2)=(4,4)$.

\begin{figure}[ht]
\centering
\renewcommand{\arraystretch}{0.8}
\begin{minipage}[t]{0.435\textwidth}
\centering
{\footnotesize Before refinement}\\[.2em]
\begin{tabular}{r|cccc|l}
\toprule
 &   $1$ & $2$  & $3$ & $4$ & $p_1$ \\ 
\midrule
{$1$} & $1/4$ & $1/4$ & $0$ & $0$ & {$1/2$}\\
$2$ & $0$ & $0$ & $1/4$ & $0$ & {$1/4$}\\
\rowcolor{highlightrow}
$3$ & $0$ & $0$ & $0$ & $1/4$ & {$1/4$}\\
\\
\midrule
$p_2$ & $1/4$ & $1/4$ & $1/4$ & $1/4$ & \\
\bottomrule
\end{tabular}
\end{minipage}
\hfill
\raisebox{-2.4cm}{ \Large $\Longrightarrow$}
\hfill
\begin{minipage}[t]{0.435\textwidth}
\centering
{\footnotesize After refinement}\\[.2em]
\begin{tabular}{r|cccc|l}
\toprule
 & $1$ & $2$ & $3$ & $4$ & $p_1'$ \\
\midrule
{$1$} & $1/4$ & $1/4$ & $0$ & $0$ & {$1/2$}\\
$2$ & $0$ & $0$ & $1/4$ & $0$ & {$1/4$}\\
\rowcolor{highlightrow}
$3$ & $0$ & $0$ & $0$ & $1/8$ & {$1/8$}\\
\rowcolor{highlightrow}
$4$ & $0$ & $0$ & $0$ & $1/8$ & {$1/8$}\\
\midrule
$p_2$ & $1/4$ & $1/4$ & $1/4$ & $1/4$ & \\
\bottomrule
\end{tabular}
\end{minipage}
\caption{\footnotesize The Fr\'echet--Hoeffding upper bound $M$ is applied to the same data. Refining category $3$ into categories $3$ and $4$, holding $p_2$ fixed, leaves rows $1$ and $2$ completely unchanged and splits row $3$'s mass in exact proportion to the new categories' shares.}
\label{fig:mincopula}
\end{figure}

To find $f(p)(2,3)$ in the left panel, observe that the cumulative probability of $(2,3)$ is $3/4$; subtracting the cumulative probabilities of $(2,2)$ ($1/2$) and $(1,3)$ ($1/2$), and adding back the cumulative probability of $(1,2)$ ($1/2$), gives $f(p)(2,3)=3/4-1/2-1/2+1/2=1/4$. Every other entry follows the same computation. This model always recommends a joint distribution whose support is weakly ordered: here, the support is concentrated on $(1,1)\leq(1,2)\leq(2,3)\leq(3,4)$, with all other cells receiving zero probability.

Unlike the maximal coupling of Figure~\ref{fig:maxcoupling}, refining category $3$ here leaves rows $1$ and $2$ completely unchanged, and simply splits row $3$'s mass in proportion to the new categories' shares. Hence, this construction satisfies IA.






\section{Main Result}\label{sec:result} We now present our main result: (i) every copula-based model satisfies the IA property, and (ii) no other models beyond these satisfy IA.

\begin{theorem}\label{th:1}A model satisfies the IA property if and only if it is a copula model.\end{theorem}

Before offering a formal proof, we will explain why this result is true.  The following sketch can be turned into a formal proof quite easily, but it is notationally cumbersome.  In fact, much of the difficulty with this result, which is conceptually quite easy, lies in the notation.

Because it is relatively simple to verify that any copula model satisfies the IA property, we will focus our effort on the converse direction, and will illustrate the result for the case $n=2$.  This sketch will never require any more than $3$ bins; and indeed, the theorem would go through with this domain restriction unchanged.  

The basic idea is to begin with a model, and define an associated function, which is then verified to be a copula.  The definition of this function is straightforward and requires only two bins for each dimension:  $\{1,2\}$.  We fix any $x\in [0,1]^2$, and define $C(x)$ as the joint probability given to the pair of bins $(1,1)$ when the underlying marginals are $(x_1,1-x_1)$ and $(x_2,1-x_2)$.  By the definition of a model, it is clear that $C(1,t)=t$ and $C(t,1)=t$ and that $C(t,0)=0$ and $C(0,t)=0$.  The only remaining thing to verify is that $C$ is actually a cumulative distribution function.  To this end, we simply need to verify that a version of the \emph{inclusion-exclusion} principle is true.  Here, we will show that if $y_1 < x_1$ and $y_2 < x_2$, then $C(x_1,x_2)-C(y_1,x_2)-C(x_1,y_2)+C(y_1,y_2)\geq 0$.  This inequality requires that the box with vertices $(y_1,y_2)$, $(x_1,y_2)$, $(y_1,x_2)$, $(x_1,x_2)$ has nonnegative measure.  In Figure~\ref{fig:proof}, we are trying to show that the measure of box $a$ is nonnegative.  If $C$ were a cumulative distribution function, we would find the measure of $a$ from $C$ in the following way:  the measure of each of the four boxes in Figure~\ref{fig:boxes} comes directly from $C$; for example, the measure of box $b$ is exactly $C(x_1,x_2)$, and the measure of box $c$ is exactly $C(x_1,y_2)$.  So, to find the measure of box $a$, we would subtract the measures of boxes $c$ and $d$ from the measure of box $b$, and add back the measure of box $e$.

Now we leverage IA.  For example, we have defined $C(x_1,x_2)$ as the application of the model to a two-bin environment, where the bins have marginals $(x_1,1-x_1)$ and $(x_2,1-x_2)$.  But consider instead a three-bin environment where the marginals are $(y_1,x_1-y_1,1-x_1)$ and $(y_2,x_2-y_2,1-x_2)$.  Then IA stipulates that $C(x_1,x_2)$ (the measure of box $b$) is also the total probability of the bins $(1,1)$, $(1,2)$, $(2,1)$ and $(2,2)$ in this three-bin environment.  All four ``sub-boxes'' of box $b$ are shaded, reflecting these four bins.  Similarly, IA tells us that the measure of $c$---which is $C(x_1,y_2)$---is the total probability of the bins $(1,1)$ and $(2,1)$.  These are the two ``sub-boxes'' of box $b$ that constitute box $c$.  And so forth.  Taken together, the application of the inclusion-exclusion principle then indicates that the implied measure of $a$ is given by the joint probability of bins $(2,2)$---which is by definition nonnegative.

\begin{figure}[h!]
\centering
\begin{tikzpicture}[scale=6]
  \fill[gray!30] (\ya,\yb) rectangle (\xa,\xb);
  \node at ({(\ya+\xa)/2},{(\yb+\xb)/2}) {$a$};
  \iagrid
  \fill (\xa,\xb) circle (0.25pt);
  \fill (\ya,\yb) circle (0.25pt);
  \node[above right] at (\xa,\xb) {$(x_1,x_2)$};
  \node[above left] at (\ya,\yb) {$(y_1,y_2)$};
\end{tikzpicture}
\caption{Proof Sketch}
\label{fig:proof}
\end{figure}
 
\begin{figure}[h!]
\centering
\iabox{\xa}{\xb}{b}\hfill
\iabox{\xa}{\yb}{c}\hfill
\iabox{\ya}{\xb}{d}\hfill
\iabox{\ya}{\yb}{e}
\caption{Continuation of Sketch}
\label{fig:boxes}
\end{figure}

\begin{proof}
We first prove the ``only if" statement. 
For $t\in [0,1]$, let $b_t$ be the binary distribution on $\{1,2\}$ given by $b_t(1)=t=1-b_t(2)$.
For $x\in [0,1]^n$, write $b_x=(b_{x_1},\dots,b_{x_n})$. 
Let $C:[0,1]^n\to [0,1]$ be given by  $C(x)=F_{f(b_x)}(1,\dots,1)$ for   $x\in [0,1]^n$.
Repeatedly applying IA, we get, for any $p\in \Delta^n_{\rm ma}$ and $s\in [z]$, 

\begin{equation}\label{eq:copula-formula}
\begin{array}{rcl}
F_{f(p)}(s)  &=&  f(b_{F_{p_1}(s_1)},\dots,b_{F_{p_n}(s_n) })(1,\dots,1)  \\
&=&C(F_{p_1}(s_1),\dots,F_{p_n}(s_n)).
\end{array}
\end{equation}

Therefore, $f$ is a copula model if we can prove that $C$ is a copula. From $C(x)=F_{f(b_x)}(1,\dots,1)$, we know that  $C(1,\dots,1,t,1,\dots,1)=t$ for any $t\in [0,1]$; this justifies that $C$ has standard uniform marginals.
It remains to show that $C$ is a distribution function. 

Let us verify that $C$ is increasing.
Fix $x\in [0,1]^n$ and $\delta\in (0,1-x_1]$. 
Take  $p\in \Delta^n_{\rm ma}$, $s\in [z]$ and $\epsilon>0$
such that $(F_{p_1}(s_1),\dots,F_{p_n}(s_n))=x $
and $ F_{p_1}(s_1+\epsilon) =x_1+\delta$.
This implies $(F_{p_1}(s_1+\epsilon),F_{p_2}(s_2),\dots,F_{p_n}(s_n))=x+(\delta,0,\dots,0)$.
By \eqref{eq:copula-formula}, we have $C(x+(\delta,0,\dots,0))\ge C(x)$ because the distribution function  $F_{f(p)}$ is increasing. Repeating this argument for the other components, we get that $C$ is increasing. 

To show that $C$ is a distribution function, take $k\in \N$ and let $G_k=\{ j/2^{k}:j\in [2^k]\}$, that is, the $(1/2^{k})$-grid.
Let $F^k$ be the distribution function on $(G_k)^n$ given by $F^k(s/2^k)=F_{f(p)}(s)$ where $p_1,\dots,p_n$ are uniform distributions on $[2^k]$; that is, we scale the distribution $F_{f(p)}$ by $1/2^k$. 
Using \eqref{eq:copula-formula}, we get, for $s\in [2^k]^n$, 
$$
C\left(\frac {s_1}{2^k},\dots,\frac{s_n}{2^k}\right) = F_{f(p)} (s) = F^k \left(\frac {s_1}{2^k},\dots,\frac{s_n}{2^k}\right).
$$
Therefore, $C$ coincides with a distribution function $F^k$ on $(G_k)^n$, and also on $(G_k\cup \{0\})^n$, since both functions vanish whenever a coordinate is zero. 
For any $x\in [0,1]^n$, let $a^k, b^k \in (G_k\cup\{0\})^n$ be 
the componentwise floor and ceiling of
$x$  with respect to the grid $(G_k\cup\{0\})^n$, respectively. We have, by the monotonicity of $C$ and $F^k$,  as $k\to\infty$,
$$|F^k(x)-C(x)| \le F^k( b^k) - C( a^k) $$ $$= F^k( b^k) - F^k( a^k)  \le  \Vert b^k-a^k \Vert \le n2^{-k}\to 0,$$
where  $\Vert\cdot \Vert$ is the $1$-norm and the penultimate inequality follows from the fact that the marginals of $F^k$ are uniform on $G_k$.
Hence, $F^k\to C$ pointwise as $k\to \infty$. 
Because the support of $F^k$ is contained in $[0,1]^n$, and since $[0,1]^n$ is compact metric, Theorem 15.11 of \citet{aliprantis2006infinite} establishes that the set of probability measures on $[0,1]^n$ endowed with its weak topology is also compact and metrizable.  Take a weakly convergent subsequence of $F^k$ and denote its limit by $F$.  The marginals of the limit are uniform on $[0,1]$, since the marginals of $F^k$ are uniform on $G_k$.  Thus, $F$ is continuous, so the cdfs along this subsequence converge to $F$ everywhere.  Since they also converge pointwise to $C$, we also know $F=C$.  Therefore, $C$ is a distribution function with uniform marginals, and thus $C$ is a copula. 

Next we prove the ``if'' statement. Suppose that $f$ is a copula model for  copula $C$.  
Let $P_C$ be  the probability measure with distribution function $C$. For $p\in \Delta_{\rm ma}^n$ and $s\in[z]$,
$$
f(p)(s) =  P_C\left(\bigtimes_{k\in N} [F_{p_k}(s_k-1), F_{p_k}(s_k)] \right).
$$
For any $i\in [n]$ and $j$, we have
$$
f(T_i^j(p))(s) = P_C\left( \bigtimes_{k \in N } \left[\hat F_{p_k}(s_k-1), \hat F_{p_k}(s_k) \right]     \right),
$$
where $\hat F_{p_k}=F_{p_k}$ for $k\ne i$ and $\hat F_{p_k}=F_{p^j_i}$ if $k=i$.
Note that $  [F_{ p^j_i}(s_i-1), F_{p^j_i}(s_i)]$ is equal to $[F_{ p_i}(s_i-1), F_{p_i}(s_i)]$
if $s_i<j$,    to $[F_{ p_i}(s_i-1), F_{p_i}(s_i+1)]$ 
if $s_i=j$, and   to $ [F_{ p_i}(s_i), F_{p_i}(s_i+1)]$ if $s_i>j$.
This shows \eqref{eq:condition-main}, and hence IA holds.  
\hfill $\square$ \end{proof}

Theorem~\ref{th:1} asserts that the only models satisfying IA are the copula models.  Three technical remarks are in order.

\begin{itemize}
\item Our notion of a model allows any $z\in\mathbb{N}^N$.  However, we could restrict the domain of $z$ to any $\times_i[1,\hat{z}_i]$, where $\hat{z}_i\in \mathbb{N}\cup \{+\infty\}$.  So long as $\hat{z}_i\geq 3$, the characterization still holds (a different, more tedious proof is necessary, however).

\item We assume that $f$ depends on $\mathbf{z}$, but is otherwise independent of the names of the ordered objects in $[z_i]$.  This means that our approach is implicitly ordinal, but also means our approach is highly flexible.  From a practical standpoint, the underlying map carrying ``real world data'' to the members of $[z_i]$ can be ignored in applications.  However, we can envision a more general approach, whereby $f$ is permitted to depend on the names of the elements of $[z_i]$.  For example, consider the ``low, medium,  high'' example described in the introduction.  $L<M<H$ constitutes a three-alternative ordered set, and is modeled by $z_i = 3$.  If low and medium are merged, we get $LM<H$, a two-element set modeled by $z_i=2$; similarly, when medium and high are merged, we get $L<MH$.  Our framework applies the same function $f$ to the two distinct merged environments.  Instead, we could allow $f$ to depend explicitly on the names of the alternatives, rather than on $z_i$.  The most natural way to do this would be to assume ambient totally ordered sets $X_i$ for each $i\in [n]$ ($X_i =\mathbb{R}$ seems a natural example), and for each $i$, a finite ordered partition of $X_i$, say $\pi_i$; the set of $\pi_i$ could be restricted to some class.  In our running example, $X_i=\{L,M,H\}$ and when $L$ and $M$ are merged, we would consider $\pi_i =\{\{L,M\},\{H\}\}$.  In our current framework, $z_i=|\pi_i|$, the cardinality of the ordered partition, but we could allow $f$ to depend on $\bigtimes_i \pi_i$ instead of only on the cardinality of each $\pi_i$.  An IA property is easy to define for this environment.  While we leave the study of this property to future research, we conjecture that it relates to a parametric generalization of the notion of copula.

\item Models satisfying IA allow us to consider the introduction of a new dimension.  If $n$ is very large, then at any given time we may consider only some subset $M\subseteq N$, with the marginal distribution of each variable indexed by $M\subseteq N$ being a point mass.   However, even if we do not wish to bound $n$, the introduction of a new dimension, say, $n+1$, poses no difficulty.  Copulas can always be extended to include another dimension.  Even more, we are free to work with $[0,1]^{\mathbb{N}}$, endowed with the Borel product $\sigma$-algebra.  We could define a copula in this case to be any measure $\mu$ with uniform marginals, and consider at any point in time only finite subsets of $\mathbb{N}$ with the induced marginal cdf on those elements.  Such probability measures $\mu$ obviously exist, owing to the Kolmogorov extension theorem, for example a joint distribution of an independent infinite sequence of uniform random variables on $[0,1]$.  Symmetric ones are easy to describe; for example analogues of the product copula and Fr\'echet upper bound are natural candidates. We investigate these more below.

\end{itemize}

\section{{Unordered} Categorical Variables}\label{sec:cat}

The preceding results dealt with the case in which choices came from some ordered set.  In this section, we investigate categorical data, where observations need not be ordered.  A familiar example of such choices might be transportation choices:  observations consist of a choice between bus and train.  Now, suppose we can refine choices further, so that we can observe whether a bus is red or blue.  There is no natural order across red bus, blue bus, and train.  Because there is no natural notion of ordering here, there is no natural notion of what it means for two choices to be adjacent.

We can discuss this issue in our setup.  Up to now, IA allows us to combine two adjacent bins (represented by integers $j$ and $j+1$) in a marginal distribution.  
If two non-adjacent bins $j$ and $k$ are combined, it is not immediately clear whether we should place this new bin at the location of $j$ or $k$. Indeed, to allow for combining arbitrary bins for a probability $q\in \Delta([w])$, the interpretation of location among the points in $[w]$ is lost. {Moreover, without an ordinal structure, adjacency cannot be properly discussed.}

Inspired by this observation, we propose the following symmetry property that removes the location from points in $[w]$. 
Let $\Sigma_w$ be the set of all permutations on $[w]$ for $w\in \N$. 
For $q\in \Delta([w])$ and $\sigma \in \Sigma_w$, $q^{\sigma}(s)=q(\sigma(s))$ for all $s\in [w]$. Let $T_i^\sigma$ be the mapping that maps $p$ to 
$(p_1,\dots,p_{i-1}, p_i^\sigma,p_{i+1},\dots, p_n)$, for all $i\in [n]$ and $\sigma\in \Sigma_{z_i}$.
When $T_i^\sigma$ is applied to a vector $s$, 
it means $T_i^{\sigma}(s)=(s_1,\dots,s_{i-1},\sigma(s_i),s_{i+1},\dots,s_n)$.

Let $M=[m]$ be a subset of $N$.  In the following definitions, $M$ should be understood as the set of categorical choices, whereas the remaining members of $N$ are to be understood as ordinal.  So the following definitions allow for a combination of categorical as well as ordinal data.

\begin{enumerate} 
 \item [] \textbf{$M$-neutrality:}  
For $p\in \Delta_{\rm ma}^n$, $i\in M$, $\sigma\in \Sigma_{z_i}$ and $s\in [z]$, 
$f(T^\sigma_i(p))(s)= f(p)(T^{\sigma}_i(s)). $  
\end{enumerate}
 
The interpretation of $M$-neutrality is that switching the labels of two bins (or reshuffling them among several bins) within $M$ does not change the joint distribution up to the corresponding relabeling.  Intuitively, the names of the bins for variables in $M$ do not matter (the standard neutrality notion of social choice).

\begin{theorem}
\label{th:m-neu} 
A model satisfies IA and $M$-neutrality  if and only if it is a copula model with $C(x)=(\prod_{i\in M} x_i )\tilde C((x_j)_{j\in N\setminus M})$, where $\tilde C$ is a copula on $[0,1]^{n-m}$, with the convention $\tilde C \equiv 1$ if $m=n$.
\end{theorem}

\begin{proof} 
Let $P_C$ be the probability measure with distribution function $C$. If $f$ is the copula model with copula $C$ of the stated form, then IA follows from Theorem \ref{th:1}, and for $i\in M$ we have $$f(T^\sigma_i(p))(s)= p_i(\sigma(s_i) )  {P_C(s_{-i})} = f(p)(T^{\sigma}_i(s)),$$ where $P_C(s_{-i})$ is the probability of the set $\{u\in[0,1]^n:F_{p_j}(s_j-1)<u_j\leq F_{p_j}(s_j),\mbox{ }j\neq i\}$ under $P_C$. Hence, $M$-neutrality  holds.  

Next, we show that if a copula model with copula $C$ satisfies $M$-neutrality, then $C$ must be the copula in the theorem. Let $p_1,\dots,p_n$ be uniform distributions on $[k]$ for $k\in \N$. For $s\in [k]^n$, let $R_s = \bigtimes_{j\in N}((s_j-1)/k,s_j/k]$.  Then $f(p)(s)=P_C(R_s)$.  Note that $T^\sigma_i(p)=p$  for each $i\in N$. Hence, for $i\in M$, we have $$P_C(R_s)=P_C(R_{T_i^{\sigma}(s)}) $$ for all $\sigma \in \Sigma_k$. Repeatedly applying this equality, we get that,  for each given $(s_i)_{i\in N\setminus M}$, $P_C$ assigns equal probability to all $k^m$ rectangles $R_s$ obtained by varying $(s_i)_{i\in M}\in [k]^m$. Thus each such rectangle has $1/k^m$ times the probability of the event imposing only the coordinates outside $M$.  Summing over rectangles establishes that the $M$-marginal distribution is uniform and independent of the remaining coordinates at all rational grid cutoffs.  Since $k$ is arbitrary and copulas are continuous, the same holds at all cutoffs in $[0,1]^n$.  Hence, the copula has the form $C(x)=(\prod_{i\in M} x_i )\tilde C((x_j)_{j\in N\setminus M})$, where $\tilde C$ is the marginal copula on the coordinates in $N\setminus M$.
\hfill $\square$ \end{proof}

 If all variables are categorical, and we insist on the neutrality property, only the independence model can be used.  On the other hand, we have some freedom in the choice of our copula for ordinal variables; for categorical variables, the model behaves as the independence model, but otherwise is completely flexible.

\section{Other Applications}\label{sec:add}

\subsection{Data Preservation}

In this section, we present two further corollaries of our main result. The first concerns an environment in which each dimension has the same number of bins and the observed marginal distributions are identical. In line with the ideas behind progressive random choice (PRC), we could hypothesize that randomness in this setting stems from variation in a single underlying index. Under this interpretation, the data should perfectly reveal that index whenever the observed marginals coincide.

An alternative motivation arises when the model is used to predict joint demand under shared capacity constraints. For a simple example, suppose consumers order two types of goods, to be delivered by the postal system:  medications and groceries. A planner seeking to ensure that all orders can be delivered, even in the worst case, would consider a scenario in which demand for both goods is high among exactly the same households or at exactly the same times. The following axiom models this ``worst case'' in the simple environment in which marginal demand for all goods exactly coincides.

In the following, $\delta_{(s,\ldots, s)}$ refers to a point mass on $(s,\ldots,s)$.

\begin{enumerate}
 \item [] \textbf{Data preservation:}
  Let $q\in\Delta([k])$ and let $p\in\Delta^n(k,\ldots,k)$ satisfy for all $i$, $p_i=q$.  Then $f(p)=\sum_s q(s)\delta_{(s,\ldots,s)}$.
\end{enumerate}

\begin{proposition}A model $f$ satisfies IA and data preservation if and only if it is a copula model with $C(x)=\min_i x_i$.\end{proposition}

\begin{proof}That the upper bound satisfies data preservation is obvious.  Conversely, and using the notation of the proof of Theorem~\ref{th:1}, for any $x\in [0,1]$, by data preservation, $f(b_x,\ldots,b_x)(1,\ldots,1)=x$.  So, $C(x,\ldots,x)=x$.  Let $y\in[0,1]^n$ and suppose that $y_i \leq y_j$ for all $j$. Then $y_i = C(y_i,\ldots,y_i)\leq C(y)\leq C(\mathbf{1}-e^i(1-y_i))=y_i$, where $\mathbf{1}$ is the all-ones vector.   Conclude that $C(y)=\min_i y_i$.\hfill $\square$ \end{proof}

This proposition shows that, under the standing assumptions of our main result, a requirement imposed only on identical marginals uniquely determines a specific model. Data preservation requires identical marginal distributions to translate into perfect agreement across dimensions, and the proposition establishes that this requirement selects the comonotonic copula, $C(x)=\min_i x_i$. The result thus provides an axiomatic foundation for modeling dependence through a single source of randomness. In the demand interpretation, it formalizes the concern that high demand for different goods occurs together, placing simultaneous pressure on shared capacity.

\subsection{Symmetric Extendability}

Here, we are interested in an environment in which the planner or economist is agnostic as to the nature of the goods in question.  To this end, we require that the names of those goods are irrelevant.  To begin, let $\sigma:[n]\rightarrow[n]$ denote a permutation.

\begin{itemize}
\item For $z\in\mathbb N^n$, define $z\circ\sigma\in\mathbb N^n$ by
$(z\circ\sigma)_i=z_{\sigma(i)}$.
\item For $p\in\Delta^n(z)$, define $p\circ\sigma\in\Delta_{ma}^n$ by
$(p\circ\sigma)_i=p_{\sigma(i)}$.  
\item For $s\in[z]$, define $s\circ\sigma\in [z\circ \sigma]$ by $(s\circ\sigma)_i=s_{\sigma(i)}$.
\end{itemize}

\begin{enumerate}
 \item [] \textbf{Symmetry:}
For every $p\in\Delta^n_{\rm ma}$, every
permutation $\sigma$ and every $s\in[z]$, $f(p\circ\sigma)\,(s\circ\sigma)=f(p)(s)$.

\end{enumerate}

The following Proposition is straightforward.
\begin{proposition}\label{pr:sym}
A model satisfies IA and symmetry if and only if it is a copula model with symmetric $C$; that is, for all $x\in [0,1]^n$ and all permutations 
$\sigma:[n]\rightarrow[n]$, $C(x)=C(x\circ\sigma)$, where $(x\circ\sigma)_i=x_{\sigma(i)}$.
\end{proposition}

Here we are not interested so much in Proposition~\ref{pr:sym} in itself, but in the idea of \emph{extendability}.  When new goods become available, what should we do?  We have mentioned already that every model satisfying IA (every copula model) is easily extendible when new goods become available.  For example, suppose that $f$ is a copula model on $n$ goods, represented by copula $C:[0,1]^n\rightarrow [0,1]$.  The function $\hat{C}(x_1,\ldots,x_n,x_{n+1})=x_{n+1}C(x_1,\ldots,x_n)$ is easily seen to be a copula as well.  So there is no issue whatsoever in introducing new goods.

However, suppose we not only want to extend the model so that it satisfies IA, but we want to extend it so that the extension satisfies both IA and symmetry.  It quickly becomes apparent that this is not always possible.  Consider the copula model represented by the symmetric copula given by the Fr\'echet-Hoeffding lower bound: $C(x_1,x_2)=\max\{0,x_1+x_2-1\}$.\footnote{More generally, there is a Fr\'echet-Hoeffding lower bound in higher dimensions, but it is not a copula.}  It turns out that there is no symmetric copula $\hat{C}(x_1,x_2,x_3)$ whose marginals on all two-dimensional subsets coincide with the Fr\'echet-Hoeffding lower bound:  observe that the distribution of the Fr\'echet-Hoeffding lower bound concentrates support on $\{(x_1,x_2):x_1+x_2=1\}$, so if there were such a $\hat{C}$, it would need to concentrate support on the points $(x_1,x_2,x_3)$ where $x_1+x_2=x_1+x_3=x_2+x_3=1$. There is a unique such point:  $(x_1,x_2,x_3)=(1/2,1/2,1/2)$, and so $\hat{C}$ must be the point mass on $(1/2,1/2,1/2)$, which does not have the lower bound as a marginal.\footnote{This is a general phenomenon.  See \citet{lauzier2023} for a technical treatment.}

Suppose we are given a model $f:\Delta^n_{\rm ma}\rightarrow\Delta^n_{\rm jo}$.  Let $m>n$.  The next axiom takes as given a model operating on $n$ commodities, and formalizes the idea that there is a model, operating on $m$ commodities, which extends the original model.
\begin{enumerate}
 \item [] \textbf{Symmetric $m$-extendability:}
There is a symmetric model satisfying IA $\hat{f}^m:\Delta^m_{\rm ma}\rightarrow \Delta^m_{\rm jo}$ for which, for every $p\in\Delta^m_{\rm ma}$, $[\hat{f}^m(p)]_{(1,\ldots,n)}=f(p_1,\ldots,p_n)$.

\end{enumerate}

For the next result, which is basically a restatement of de Finetti's theorem, we need a few more definitions.  Denote the set of Borel probability measures on $[0,1]$ by $\mathcal{P}([0,1])$.  We endow $\mathcal{P}([0,1])$ with the weak*-topology, so that $\nu_k\rightarrow \nu^*$ if and only if for every continuous $g:[0,1]\rightarrow\mathbb{R}$ we have $\int g d\nu_k \rightarrow \int g d\nu^*$.  Then we can let $\mathcal{B}$ be the Borel sets associated with the weak* topology, and it is meaningful to speak of probability measures on $(\mathcal{P}([0,1]),\mathcal{B})$.  

Identify each $\nu\in\mathcal{P}([0,1])$ with its cdf $F_\nu(x)=\nu([0,x])$.  

\begin{theorem}An IA model $f:\Delta^n_{\rm ma}\rightarrow\Delta^n_{\rm jo}$ satisfies symmetric $m$-extendability for all $m>n$ if and only if it is a copula model for which there is a probability $\mu$ on $(\mathcal{P}([0,1]),\mathcal{B})$ for which for every $x\in [0,1]$, $x=\int F_{\nu}(x)d\mu(\nu)$ such that $C(x_1,\ldots,x_n)=\int_{\mathcal{P}([0,1])}\prod_{i=1}^nF_{\nu}(x_i)d\mu(\nu).$ \end{theorem}

\begin{proof} That any such model is $m$-extendable for every $m>n$ is simple to see:  for any $m>n$, let $\hat{C}(x_1,\ldots,x_m)=\int_{\mathcal{P}([0,1])}\prod_{i=1}^mF_{\nu}(x_i)d\mu(\nu)$; this provides a copula model generating the symmetric $m$-extension.

Conversely, let $C$ be the copula representing $f$.  $C$ is the cdf for a symmetric probability $\lambda\in\mathcal{P}([0,1]^n)$.  By Theorem~\ref{th:1} and Proposition~\ref{pr:sym}, $\hat{f}^m$ is a copula model with a symmetric copula $C^m$, which clearly satisfies $C^m(x_1,\ldots,x_n,1,\ldots,1)=C(x_1,\ldots,x_n)$.  Let $\lambda^m\in\mathcal{P}([0,1]^m)$ be the associated probability distribution.

Consider the set $[0,1]^{\mathbb{N}}$, endowed with the smallest $\sigma$-algebra for which all sets of the form $\prod_{i=1}^k B_i \times \prod_{i=k+1}^{\infty}[0,1]$ are measurable, where each $B_i\subseteq [0,1]$ is Borel-measurable.  This is the product $\sigma$-algebra, and it coincides with the $\sigma$-algebra generated by the product topology on $[0,1]^{\mathbb{N}}$; see \cite[Lemma 6.4.2(ii)]{bogachev2007}.  We will therefore call it $\sigma([0,1]^{\mathbb{N}})$ and probability measures on this set with its product $\sigma$-algebra $\mathcal{P}([0,1]^{\mathbb{N}})$.

Now, for every $m>n$ the measure $\hat{\lambda}^m\in\mathcal{P}([0,1]^{\mathbb{N}})$ defined by $\hat{\lambda}^m=\lambda^m\times \delta_{(1,1,\ldots)}$ satisfies $[\hat{\lambda}^m]_{1,\ldots,m}=\lambda^m$.  In particular, it has uniform marginals on the first $m$ coordinates.

Finally, by Tychonoff's Theorem \cite[Theorem 2.61]{aliprantis2006infinite}, since $[0,1]^{\mathbb{N}}$ is compact and metrizable \cite[Theorem 3.36]{aliprantis2006infinite}, we conclude that $\mathcal{P}([0,1]^{\mathbb{N}})$ is also compact metrizable \cite[Theorem 15.11]{aliprantis2006infinite}.  We let $\lambda^*\in\mathcal{P}([0,1]^{\mathbb{N}})$ be a limit point of the sequence $\hat{\lambda}^m$.  

Now, let $\sigma:\mathbb{N}\rightarrow\mathbb{N}$ be a finite permutation; that is, so that there is $r$ for which for all $l>r$, $\sigma(l)=l$.  Define $T_\sigma:[0,1]^{\mathbb{N}}\rightarrow[0,1]^{\mathbb{N}}$ by $T_\sigma(u)_i=u_{\sigma(i)}$, which is continuous in the product topology. 

For this $\sigma$, let $l>r$; then we know that the pushforward measure $\hat{\lambda}_{\sigma}^l$, defined as $\hat{\lambda}^l_\sigma(E)=\hat{\lambda}^l(T_{\sigma}^{-1}(E))$ is actually equal to $\hat{\lambda}^l$, since ${\sigma}$ only permutes indices $\leq r$.  This follows as $C^l$ is a symmetric copula, inducing a symmetric measure.

Since $T_\sigma$ is continuous, for every continuous real function $g$ on $[0,1]^{\mathbb{N}}$, by following a subsequence $l_j$ so that $\lim_j \hat{\lambda}^{l_j}=\lambda^*$
\[
 \int_{[0,1]^{\mathbb{N}}} g\circ T_\sigma d\lambda^*
 =\lim_{j\to\infty}\int_{[0,1]^{\mathbb{N}}} g\circ T_\sigma d \hat{\lambda}^{l_j}
 =\lim_{j\to\infty}\int_{[0,1]^{\mathbb{N}}} gd \hat{\lambda}^{l_j}
 =\int_{[0,1]^{\mathbb{N}} }gd \lambda^*.
\]
where the first equality follows from continuity of $g\circ T_{\sigma}$ and definition of weak* convergence, the second by the symmetry property, the third again by weak* convergence.

As a consequence, the probability $\lambda^*\in\mathcal{P}([0,1]^{\mathbb{N}})$ is exchangeable. Moreover, each coordinate $i$ is uniformly distributed under $\hat{\lambda}^{l}$ whenever $l\geq i$, and for all $l>n$, $[\hat{\lambda}^l]_{1,\ldots,n}=\lambda$ owing to the fact that $C^l$ extends $C$. Since the coordinate projections are continuous, these properties pass to the limit: every one-dimensional marginal of $\lambda^*$ is uniform, and $[\lambda^*]_{1,\ldots,n}=\lambda$, the probability represented by the copula $C$.

Then we may conclude by de Finetti's Theorem (see for example \citet{hewitt1955} Theorem 7.2, they state their result for Baire $\sigma$-algebras, but the relevant spaces here are compact metric, so their Baire and Borel $\sigma$-algebras coincide, \cite[Lemma 4.65(2)]{aliprantis2006infinite} that there is a probability measure $\mu$ on $\mathcal{P}([0,1])$ such that $\lambda^*=\int \nu^{\mathbb{N}}d\mu(\nu)$, where $\nu^{\mathbb{N}}$ denotes the unique measure on $[0,1]^{\mathbb{N}}$ whose marginals on all finite cylinders of dimension $l$ coincide with the product of $\nu$ with itself $l$ times.

This tells us directly that $C(x_1,\ldots,x_n)=\int \prod_{i=1}^n F_{\nu}(x_i)d\mu(\nu)$.  The fact that $\int F_{\nu}(x)d\mu(\nu)=x$ owes to the fact that the marginals of $\lambda^*$ on any single dimension are uniform.  \hfill $\square$ \end{proof}

Of note is that the ``mixing measure'' for the Fr\'echet Upper Bound, where $\mu$ is the uniform distribution over the $\delta_x$ point masses on $x$, is a ``mean-preserving spread'' of every other distribution over $\mathcal{P}([0,1])$ with uniform barycenter, while every such distribution is a mean-preserving spread of the point mass on the uniform (the product measure). There does not seem to be a particularly interesting way of studying this comparative static in terms of the models themselves.

\begin{figure}[h!]
\centering
\includegraphics[width=\textwidth]{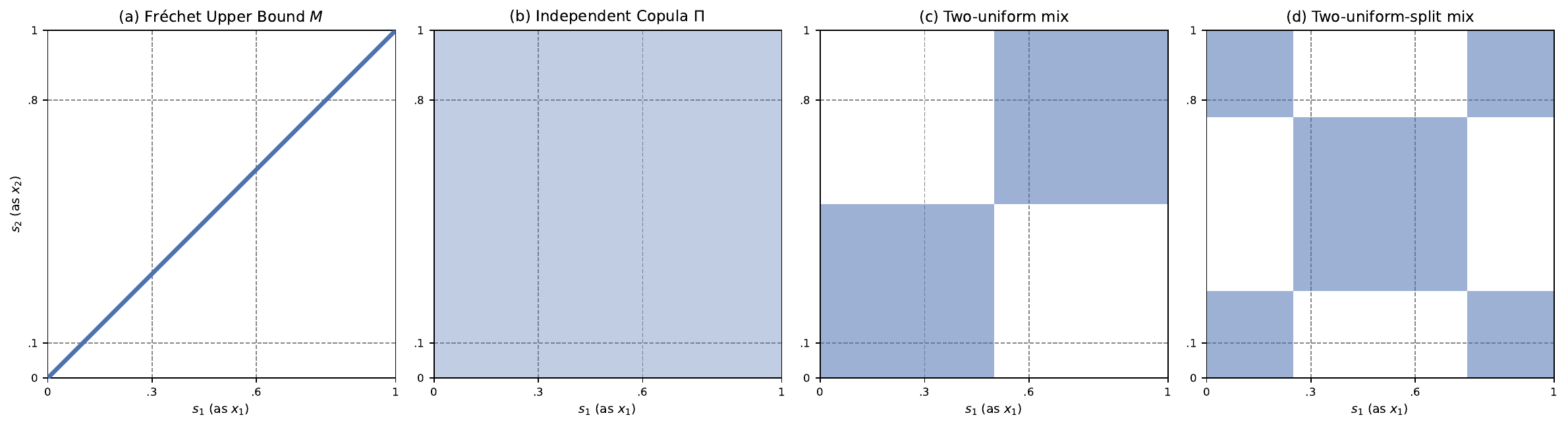}
\caption{\footnotesize The support of each associated copula distribution in the $(x_1,x_2)$ unit square, with the marginal breakpoints of $p_1=(.3,.3,.4)$ and $p_2=(.1,.7,.2)$ overlaid as dashed lines. Panel (a) concentrates on the diagonal, panel (b) spreads uniformly, and panels (c) and (d) each split mass among a small number of blocks with no mass between them. Computing the probability assigned to each grid cell recovers the corresponding entry of Figure~\ref{fig:definetti} exactly.}
\label{fig:copula-support}
\end{figure}

Figure~\ref{fig:copula-support} illustrates four copulas. The first is the Fr\'echet-Hoeffding Upper Bound, given by a $\mu$ which is uniform on the point masses $\delta_x$. The second is the independence copula, which is a point mass on the uniform on $[0,1]$.  The third one  is a $1/2$ probability on the uniform on $[0,1/2]$, and  $1/2$ on the uniform on $[1/2,1]$, while the last one depicts  $1/2$ on $[0,1/4]\cup[3/4,1]$ and $[1/4,3/4]$. Because $C(x_1,x_2)=\int F_\nu(x_1)F_\nu(x_2)\,d\mu(\nu)$ represents two independent draws conditional on the same realized type $\nu$, all dependence between the two commodities is generated entirely by the shared randomness in which $\nu$ is drawn---not by any comonotonicity within a type. This is visible directly in the support: when $\mu$ places weight on a type $\nu$, the resulting support is the product of $\nu$'s own support with itself, which is a rectangle when the support of $\nu$ is an interval.  For the Fr\'echet upper bound (panel (a)), $\mu$ is a point mass on each degenerate $\delta_x$, so every such rectangle collapses to a single point on the diagonal $x_1=x_2$, and the aggregate support is the familiar comonotone line. For the product measure (panel (b)), $\mu$ is a point mass on the uniform distribution itself, so the single rectangle is the entire square. Panels (c) and (d) are genuine two-point mixtures, and each contributes two or more disjoint rectangles---the population divides into a small number of types, each internally independent, with no mass connecting one type's outcomes to another's.

The dashed lines mark the breakpoints of $p_1=(.3,.3,.4)$ and $p_2=(.1,.7,.2)$. Computing the probability assigned to each resulting grid cell recovers the corresponding entry of Figure~\ref{fig:definetti} exactly: for instance, cell $(1,1)$ in every panel is precisely the mass of the shaded region lying inside $[0,.3]\times[0,.1]$. Figure~\ref{fig:definetti} can thus be read as a discretization of Figure~\ref{fig:copula-support} along this particular grid, and different choices of grid would discretize the same four continuous distributions into different tables.

Figure~\ref{fig:definetti} shows four static two-good distributions, but the content of the extendability theorem is not really about describing such distributions---it is about which of them can be \emph{grown}, coordinate by coordinate, without ever breaking symmetry or IA. Figure~\ref{fig:definetti3} makes that content concrete by extending to a third commodity with $p_3=(.4,.6)$, using the same mixing measure $\mu$ in each case and the formula $\hat{C}(x_1,x_2,x_3)=\int_{\mathcal{P}([0,1])}F_\nu(x_1)F_\nu(x_2)F_\nu(x_3)\,d\mu(\nu)$.

\begin{figure}[ht]
\centering
\small

\begin{minipage}[t]{0.44\textwidth}
\centering
 (a)  Fr\'echet upper bound \\[.4em]
{\footnotesize \begin{tabular}{c|CCC|S}
\toprule
$s_1\backslash s_2$ & $1$ & $2$ & $3$ & $p_1$ \\
\midrule
$1$\rs & $.1$ & $.2$ & $0$ & $.3$ \\
$2$\rs & $0$ & $.3$ & $0$  & $.3$ \\
$3$\rs & $0$ & $.2$ & $.2$  & $.4$ \\
\midrule
$p_2$\rs & $.1$ & $.7$ & $.2$ &  \\
\bottomrule
\end{tabular}}
\end{minipage}
\hfill
\begin{minipage}[t]{0.44\textwidth}
\centering
 (b) Independent Copula \\[.4em]
{\footnotesize \begin{tabular}{c|CCC|S}
\toprule
$s_1\backslash s_2$ & $1$ & $2$ & $3$ & $p_1$ \\
\midrule
$1$\rs & $.03$ & $.21$ & $.06$ & $.3$ \\
$2$\rs & $.03$ & $.21$ & $.06$ & $.3$ \\
$3$\rs & $.04$ & $.28$ & $.08$ & $.4$ \\
\midrule
$p_2$\rs & $.1$ & $.7$ & $.2$ & \\
\bottomrule
\end{tabular}}
\end{minipage}

\vspace{1.0em}

\begin{minipage}[t]{0.44\textwidth}
\centering
 (c) Two-uniform mix \\[.4em]
{\footnotesize \begin{tabular}{c|CCC|S}
\toprule
{ $s_1\backslash s_2$} & $1$ & $2$ & $3$ & $p_1$ \\
\midrule
$1$\rs & $.06$ & $.24$ & $0$ & $.3$ \\
$2$\rs & $.04$ & $.22$ & $.04$ & $.3$ \\
$3$\rs & $0$ & $.24$ & $.16$ & $.4$ \\
\midrule
$p_2$\rs & $.1$ & $.7$ & $.2$ & \\
\bottomrule
\end{tabular}}
\end{minipage}
\hfill
\begin{minipage}[t]{0.44\textwidth}
\centering
 (d) Two-uniform-split mix \\[.4em]
{\footnotesize \begin{tabular}{c|CCC|S}
\toprule
$s_1\backslash s_2$ & $1$ & $2$ & $3$ & $p_1$ \\
\midrule
$1$\rs & $.05$ & $.15$ & $.1$ & $.3$ \\
$2$\rs & $0$ & $.3$ & $0$ & $.3$ \\
$3$\rs & $.05$ & $.25$ & $.1$ & $.4$ \\
\midrule
$p_2$\rs & $.1$ & $.7$ & $.2$ & \\
\bottomrule
\end{tabular}}
\end{minipage}
\normalsize
\caption{\footnotesize  Four representations of the same data $p_1=(.3,.3,.4)$, $p_2=(.1,.7,.2)$, each generated by a different mixing measure $\mu$ via $C(x_1,x_2)=\int F_\nu(x_1)F_\nu(x_2)\,d\mu(\nu)$. Panel (a) takes $\mu$ uniform over the point masses $\delta_x$, $x\in[0,1]$, recovering the Fr\'echet upper bound $M$; panel (b) takes $\mu$ a point mass on the uniform distribution, recovering the independent copula $\Pi$. Panels (c) and (d) are genuine two-point mixtures---equal weight on $U[0,\tfrac12]$ and $U[\tfrac12,1]$ in (c), and on $U[0,\tfrac14]\cup[\tfrac34,1]$ and $U[\tfrac14,\tfrac34]$ in (d)---interpolating between these two extremes. Rows index $s_1$ and columns index $s_2$; in every panel the margins recover $p_1$ and $p_2$ exactly. }
\label{fig:definetti}
\end{figure}

\begin{figure}[ht]
\centering
\small

\begin{minipage}[t]{0.44\textwidth}
\centering
 (a)  Fr\'echet Upper Bound \\[.4em]
{\footnotesize \begin{tabular}{c|CCC|S}
\toprule
$s_1\backslash s_2$ & $1$ & $2$ & $3$ & $p_1$ \\
\midrule
$1$\rs & \splitcell{.1}{0} & \splitcell{.2}{0} & \splitcell{0}{0} & $.3$ \\
$2$\rs & \splitcell{0}{0} & \splitcell{.1}{.2} & \splitcell{0}{0} & $.3$ \\
$3$\rs & \splitcell{0}{0} & \splitcell{0}{.2} & \splitcell{0}{.2} & $.4$ \\
\midrule
$p_2$\rs & $.1$ & $.7$ & $.2$ &  \\
\bottomrule
\end{tabular}}
\end{minipage}
\hfill
\begin{minipage}[t]{0.44\textwidth}
\centering
 (b) Independent Copula \\[.4em]
{\footnotesize \begin{tabular}{c|CCC|S}
\toprule
$s_1\backslash s_2$ & $1$ & $2$ & $3$ & $p_1$ \\
\midrule
$1$\rs & \splitcell{.012}{.018} & \splitcell{.084}{.126} & \splitcell{.024}{.036} & $.3$ \\
$2$\rs & \splitcell{.012}{.018} & \splitcell{.084}{.126} & \splitcell{.024}{.036} & $.3$ \\
$3$\rs & \splitcell{.016}{.024} & \splitcell{.112}{.168} & \splitcell{.032}{.048} & $.4$ \\
\midrule
$p_2$\rs & $.1$ & $.7$ & $.2$ & \\
\bottomrule
\end{tabular}}
\end{minipage}

\vspace{1.0em}

\begin{minipage}[t]{0.44\textwidth}
\centering
 (c) Two-uniform mix \\[.4em]
{\footnotesize \begin{tabular}{c|CCC|S}
\toprule
{ $s_1\backslash s_2$} & $1$ & $2$ & $3$ & $p_1$ \\
\midrule
$1$\rs & \splitcell{.048}{.012} & \splitcell{.192}{.048} & \splitcell{0}{0} & $.3$ \\
$2$\rs & \splitcell{.032}{.008} & \splitcell{.128}{.092} & \splitcell{0}{.04} & $.3$ \\
$3$\rs & \splitcell{0}{0} & \splitcell{0}{.24} & \splitcell{0}{.16} & $.4$ \\
\midrule
$p_2$\rs & $.1$ & $.7$ & $.2$ & \\
\bottomrule
\end{tabular}}
\end{minipage}
\hfill
\begin{minipage}[t]{0.44\textwidth}
\centering
 (d) Two-uniform-split mix \\[.4em]
{\footnotesize \begin{tabular}{c|CCC|S}
\toprule
$s_1\backslash s_2$ & $1$ & $2$ & $3$ & $p_1$ \\
\midrule
$1$\rs & \splitcell{.025}{.025} & \splitcell{.065}{.085} & \splitcell{.05}{.05} & $.3$ \\
$2$\rs & \splitcell{0}{0} & \splitcell{.09}{.21} & \splitcell{0}{0} & $.3$ \\
$3$\rs & \splitcell{.025}{.025} & \splitcell{.095}{.155} & \splitcell{.05}{.05} & $.4$ \\
\midrule
$p_2$\rs & $.1$ & $.7$ & $.2$ & \\
\bottomrule
\end{tabular}}
\end{minipage}
\normalsize
\caption{\footnotesize The extension of Figure~\ref{fig:definetti}'s four mixing measures to a third good with $p_3=(.4,.6)$. Rows index $s_1$, columns index $s_2$; each cell shows \textcolor{barone}{$s_3=1$} / \textcolor{bartwo}{$s_3=2$}. In every panel the margins recover $p_1=(.3,.3,.4)$ and $p_2=(.1,.7,.2)$ exactly.}
\label{fig:definetti3}
\end{figure}

\section{Conclusion and Related Literature}\label{sec:conclusion}

In this paper, we proposed and characterized the property of Invariant Aggregation (IA)—a condition that requires merging adjacent categories in marginal distributions not to distort the joint distribution over unaffected categories. We proved that IA holds if and only if the model is based on a copula, thereby offering a precise, axiomatic foundation for when and why copula-based modeling is appropriate in discrete choice settings. This result highlights the powerful structure of copulas in maintaining consistency under data coarsening or refinement, which is common in empirical contexts.

Our investigation was motivated by problems in stochastic choice theory, especially in contexts where joint choice data is incomplete or unobserved.   
In economics, the rationalization of data is usually motivated by an underlying behavioral model, and in some such environments, there may be either no rationalization or multiple rationalizations.  The canonical example of such a framework is the \emph{random utility model} (RUM), which we will not describe here (it is an example of a categorical model; no exogenous order on $X$ is specified); see \cite{Block,falmagne,mcfadden1978modeling,kono2023axiomatization, yildiz2016list} and \cite{dardanoni2022mixture}.

 Recent contributions focus on ordered choices.  A major contribution is due to \cite{scrum} and \cite{prc}, the so-called \emph{progressive random choice} (PRC) model; see also \cite{yildiz2024,apesteguia2023random,petri2023binary}, and \cite{masatlioglu2024growing}.  In this model, the set $X$ is endowed with a linear order and each $A_i \subseteq X$ inherits the restriction of this order.  
 \cite{chambers2025ordered} introduce copula-based models for stochastic choice and show that the progressive random choice model is a special case. Indeed, the progressive random choice model coincides with the Fr\'echet-Hoeffding upper bound model. In economics problems, we sometimes imagine new data arriving in the form of a new subset $B\subseteq X$ and a distribution over $B$.  We realized that the application of the PRC model when including $B$ in the data did not distort the marginal probabilities assigned to each member of $\bigtimes_i A_i$.  This naturally led to the investigation of the IA property here.

For future research, it is worth considering the more general notion of a model, which allows for empty or multi-valued sets of rationalizations.  It can be proved that RUM satisfies an analogue of IA tailored to such an environment, but we do not know the general class.  The class of models satisfying IA in a more general cardinally specified framework is also of interest.


\bibliographystyle{agsm}
\bibliography{choicematchingcombined}

\end{document}